\documentclass[amssymb,useAMS,prd,aps,twocolumn,superscriptaddress,preprintnumbers,nofootinbib]{revtex4-1}
\usepackage{pslatex}
\usepackage[pdftex]{graphicx}
\usepackage{psfrag}
\usepackage{epsfig}
\usepackage{color}
\usepackage{cancel}
\usepackage{slashed}
\usepackage[utf8]{inputenc}

\def\refe@jnl#1{{#1}}
\def\aj{\refe@jnl{Astron.~J.}}
\def\araa{\refe@jnl{Annu.~Rev.~Astron.~Astrophys.}}
\def\apj{\refe@jnl{Astrophys.~J.}}
\def\apjl{\refe@jnl{Astrophys.~J.~Lett.}}
\def\aap{\refe@jnl{Astron.~Astrophys.}}
\def\mnras{\refe@jnl{Mon.~Not.~R.~Astron.~Soc.}}
\def\prd{\refe@jnl{Phys.~Rev.~D}}
\def\fcp{\refe@jnl{Fund.~Cos.~Phys.}}
\def\physrep{\refe@jnl{Phys.~Rep.}}
\def\physlett{\refe@jnl{Phys.~Lett.}}

\def\invisible#1{  }

\def\lsim{\mathrel{\lower4pt\hbox{$\sim$}} 
\hskip-9.5pt\raise1.6pt\hbox{$<$}\;} 
 
\def\gsim{\mathrel{\lower4pt\hbox{$\sim$}} 
\hskip-9.5pt\raise1.6pt\hbox{$>$}\;}

\begin{document}
\preprint{ULB-TH/14-05}

\title{Can a millicharged dark matter particle emit an observable $\gamma$-ray line?}

\author{Cha\"imae El Aisati, Thomas Hambye and Tiziana Scarna}
\email{Chaimae.El.Aisati@ulb.ac.be;thambye@ulb.ac.be;tscarna@ulb.ac.be}
\affiliation{Service de Physique Th\'eorique\\
 Universit\'e Libre de Bruxelles\\ 
Boulevard du Triomphe, CP225, 1050 Brussels, Belgium}


\begin{abstract}
If a $\gamma$-ray line is observed in the near future, it will be important to determine what kind of dark matter (DM) particle could be at its origin. 
We investigate the possibility that the $\gamma$-ray line would be induced by a slow DM particle decay associated to the fact that the DM particle would not be absolutely neutral. A ``millicharge'' for the DM particle can be induced in various ways, in particular from a kinetic mixing interaction or through the Stueckelberg mechanism. We show that 
such a scenario could lead in  specific cases to an observable $\gamma$-ray line. This possibility can be considered in a systematic model-independent way, by writing down the corresponding effective theory. This allows  
for a multi-channel analysis, giving in particular upper bounds on the intensity of the associated $\gamma$-ray line from cosmic rays  emission.
Our analysis includes the possibility that in the two-body decay the photon is accompanied with a neutrino. We show that, given the stringent constraints which hold on the millicharge of the neutrinos, this is not an option, except if the DM particle mass lies in the very light KeV-MeV range, allowing for a possibility of explanation of the recently claimed, yet to be confirmed, $\sim 3.5$~KeV X-ray line.
\end{abstract}

 \maketitle
\section{Introduction}
One of the most promising ``smoking-gun'' signals for establishing the existence of the dark matter particle is the possible observation of a sharp cosmic $\gamma$-ray line from dark matter annihilation or decay \cite{Bergstrom:1988fp}.  The forthcoming Cherenkov telescopes \cite{CTA::2013}, the current Fermi large area telescope \cite{Fermi::2013} and the HESS instrument \cite{HESSII::2012} will allow to probe this possibility with further sensitivity.
If such a signal is observed in the near future, the question of the identification of the DM particle that could have caused it will become crucial.
Such a signal could be induced through annihilation, coannihilation or decay. For all these scenarios, it is generally assumed that the photon is emitted through the loop of a charged particle. Beside this general class of models, there exist other ways along which DM could emit monochromatic photons. One possibility consists in assuming that the $\gamma$-ray line is due to a $Z-Z'-\gamma$ Chern-Simons interaction \cite{Dudas:2012pb}.  
Another possibility, much less studied,
would be to consider a photon directly emitted by the DM particle. This is a priori perfectly possible if DM is not exactly neutral, but is millicharged.
For an annihilation such a possibility is not much of an option because the associated $\gamma$-ray line would be in general suppressed with respect to the total cross section, by the square of the millicharge. Given the constraints there are on the total cross section (in particular from the relic density in the thermal freezeout scenario), this would lead to a signal sizeably smaller than present or near future sensitivities. 
Instead for a decay, there is a priori more freedom because the decay lifetime is not so directly constrained by the relic density. In this work we consider such a decay possibility. 

In the following we will first consider the two main frameworks that can in a simple way justify a millicharge for the DM particle, kinetic mixing and Stueckelberg scenarios. 
In such scenarios, in order to justify that the DM particle would have a slow decay, we assume that its stability is due to an accidental symmetry that, being accidental, would be naturally broken by any UV physics. Along these lines, the decay is naturally slow because suppressed by powers of the UV scale, just as expected for the proton. The appropriate language to consider in a model-independent way the possibility of a slow decay is therefore the one of the higher-dimensional operator effective theory. Unlike for an annihilation, the use of an effective theory for a decay is fully justified since one expects a clear scale separation. Consequently, such an effective theory  allows for a systematic study of possibilities.  We will determine all dimension-five and dimension-six operators that can lead to a two-body radiative decay from a millicharged fermion, scalar or vector DM particle. These operators come in addition to the effective operators which can lead to a $\gamma$-ray line in the case where DM would be exactly neutral, given and studied in Ref.~\cite{Gustafsson::2013}. The former operators involve a covariant derivative of the millicharged field, whereas the latter ones can involve a photon only from the presence of a hypercharge or $SU(2)_L$ field strength $F_{Y,L}^{\mu\nu}$ in the operator.

In the following, we will perform a detailed analysis of the constraints that hold on the various ``millicharged operators'' for the fermionic DM case. The scalar and vector cases
will be discussed more briefly before concluding.  
A simple constraint that turns out to be relevant in some cases is that the DM particle lifetime should be larger than the age of the Universe. Another one concerns the emission of cosmic rays (CR) that could be associated to the one of the photon, either from the particle that accompanies the photon in the decay final state, or from other decays that the effective operator unavoidably predicts on top of the radiative one. Gauge invariance in particular predicts decays where the photon is replaced by a $Z$. If the electromagnetic coupling to the $Z$ is not millicharge suppressed, the flux of cosmic rays produced is much larger than the flux of monochromatic photons. 
In particular, if the particle accompanying the photon in the final state is a neutrino, which is the only Standard Model (SM) particle possibility (a decay of special interest being ``poly-monochromatic'', i.e.~monochromatic for both types of cosmic rays that are the less affected while propagating), we will see that an observable $\gamma$-ray line is not an option, unless the DM mass is quite low.
Therefore, except for this case, the possibilities we will find point towards multi-component DM scenarios.
Other constraints are related to the fact that along the Stueckelberg scenario the DM particle is charged under a new $U(1)'$ gauge group, which may be at the origin of the unsuppressed emission of the associated $Z'$.

\section{Three millicharged frameworks}
A millicharge for a particle can either be postulated as just so (from assuming an hypercharge such that $Q=T_3+Y/2$ is small) or induced from a dynamical process, typically a small mixing parameter between the SM hypercharge gauge boson and a new $U(1)'$ gauge boson.

The first option requires another particle to carry just so the same millicharge in such a way that the DM particle can decay into it.
For the more appealing second option, one can point out two simple scenarios, depending on whether the $U(1)'$ gauge boson is massless or massive.

\subsection{Massless scenario: millicharge from kinetic mixing}

In the massless case, a millicharge is induced for an originally neutral particle if the unbroken $U(1)'$ gauge boson kinetically mixes with the hypecharge gauge group \cite{Holdom::1986, Foot::1991}, 
\begin{equation}
\label{kinmix}
{\cal L} \owns -\frac{\varepsilon}{2} F_{Y\mu\nu} F'^{\mu\nu}\,.
\end{equation}
Applying first a non-unitary transformation to get rid of this non-canonical kinetic term, one can always in a second step rotate both gauge boson fields with a unitary transformation because both gauge bosons are massless. There is therefore some arbitrariness in defining both fields.
We make the convenient choice to go to the basis where the state which is essentially the hypercharge gauge boson couples to both
$Q_{SM}\equiv T_3+Y/2$ and $Q'$ generators, whereas the other one, which is essentially the $U(1)'$ gauge boson, couples only to the $Q'$ generator. It allows to put the kinetic mixing suppression in the production decay process rather than in the detection, see e.g.~Ref.~\cite{Chu::2012}.
In this basis, and after electroweak symmetry breaking, the covariant derivative $\partial_\mu+i g T_3 W_{3\mu}+i g_Y\frac{Y}{2} B_{Y\mu} +i g' Q' B'_\mu$ becomes

\begin{equation}
\setlength\arraycolsep{0.2em}
 \begin{array}{rclclcl}
D_\mu= \partial_\mu  &+& i g(T^1 W^1_\mu + T^2 W^2_\mu)\\
&+& i A^\gamma_\mu (\frac{e Q  \cos(\theta_\epsilon) }{\cos \theta_W \sqrt{1-\epsilon^2}}-\frac{g' Q' \epsilon \cos \theta_\epsilon }{\sqrt{1-\epsilon^2}})  \\
&+ & i Z_\mu (g T^3 \cos \theta_\epsilon-g_Y\frac{Y}{2} \frac{\sin \theta_\epsilon}{\sqrt{1-\epsilon^2}}+\frac{g' Q' \epsilon \sin \theta_\epsilon }{\sqrt{1-\epsilon^2}})  \\ 
&+ & i A^{ \gamma'}_\mu g' Q'\,,
\label{lagrangian_kin_mix}
\end{array}
\end{equation}
with $\tan \theta_\epsilon= \frac{\tan \theta_W}{\sqrt{1-\epsilon^2}}$. A field with charges $T_3$, $Y$, and $Q'$ couples to the photon field $A_\mu$ with charge $Q_{em}=( Q_{SM}-g' Q'  \, \epsilon/g_Y ) e'$ with $e'= g_Y  \cos \theta_\epsilon/\sqrt{1- \epsilon^2}$.  In particular, a field with $Q_{SM}=0$ acquires a millicharge $Q_{em}=-(\epsilon g' Q' /g_Y) e'$.
Note that everywhere in the following we will make the approximation $\tan \theta_\epsilon =\frac{ \tan\theta_W}{\sqrt{1-\epsilon^2}}\simeq \tan \theta_W$. 
Existing constraints on the parameters apply in general on the millicharge $Q_{em}$, rather than on $\epsilon$ directly, see below. A value of $\epsilon\simeq1$ is therefore not excluded. However, it is generally expected smaller than one. For instance, if  we consider the minimal scenario
where the  only DM couplings  are those of Eq.(\ref{lagrangian_kin_mix}), the thermal  relic abundance of the DM is provided by the annihilation into dark photons. By requiring the right dark matter abundance, we get a value for $Q'^2\alpha'= Q'^2 g'^2/4\pi$ as a function of $m_{DM}.$ This constraint together with Eq. (\ref{mc_bound}) below gives the bound $\epsilon^2 \leq 10^{-6}$, justifying our approximation.

\subsection{Massive scenario: millicharge from Stueckelberg mechanism}

It is well known that if the $U(1)'$ symmetry is spontaneously broken, so that the corresponding gauge boson becomes massive, a kinetic mixing interaction does not induce any millicharge for an originally neutral field. In the massive case there exists nevertheless the Stueckelberg option.
The Stueckelberg mechanism allows to have a massive gauge boson without breaking the corresponding gauge symmetry. We will here consider an extension of the SM by a $U(1)'$ as in \cite{Nath::2004}. This model contains a scalar which has Stueckelberg couplings to both $U(1)_Y$ and $U(1)'$. As a consequence, the neutral gauge bosons mix, and in the mass eigenstates basis  the covariant derivative reads 
\begin{equation}
\setlength\arraycolsep{0.2em}
 \begin{array}{rclclcl}
D_\mu &=& \partial_\mu  + i Z'_\mu \left( g' Q'(c_\psi c_\phi - s_\psi s_\theta s_\phi)-g T^3 c_\theta s_\psi \right. \\
&+& \left. g_Y \frac{Y}{2}(c_\psi s_\phi + s_\theta c_\phi s_\psi)\right)  \\
&+ & i Z_\mu \left(g' Q'(- s_\psi c_\phi - c_\psi s_\theta s_\phi)- g T^3 c_\theta c_\psi \right.\\ 
&+&\left. g_Y \frac{Y}{2} (- s_\psi s_\phi + s_\theta c_\phi c_\psi)\right)   \\
&+ & i A^\gamma_\mu \left(- g' Q' c_\theta s_\phi + g T^3 s_\theta +g_Y \frac{Y}{2} c_\theta c_\phi ) \right.\\
&+ & i g T^1 W^1_\mu + i g T^2  W^2_\mu\,,
\label{lagrangian_Stuck}
\end{array}
\end{equation}
where $c$ and $s$ stand for the sine and cosine of the various angles with $\tan \phi = \frac{M_2}{M_1} ,\, \tan \theta =\frac{g_Y}{g}\cos \phi,\, \tan \psi = \frac{\tan\theta \tan\phi M_{\rm W}^2}{\cos\theta(M^2_{{\rm Z}'}-(1+\tan^2 \theta)M_{\rm W}^2)}
$,
with $M_1, M_2$ the ``bare'' mass of the $U(1)', \,U(1)_Y$ gauge boson, respectively \cite{Nath::2004}.  The expression of the electromagnetic charge is
$Q_{em}= (- g'/g_Y Q' \tan\phi+Q_{SM})e'$, with $e'=g g_Y \cos\phi/\sqrt{g^2+g_Y^2 \cos^2 \phi}$. 
In this way, an originally neutral field acquires a charge of $Q_{em}= - Q' \tan \phi e'  g'/g_Y$.
Note that the Stueckelberg scenario as origin of a millicharged DM (like the  ``just-so'' scenario) might be questioned by considerations of quantum gravity/string theory \cite{Shiu::2013}.
\section{Possible two-body radiative  decays and list of effective operators that can induce them}
\label{section::effective op} 

The list of possible radiative decays that could be generated by the millicharge of a particle is extremely reduced and in this sense points towards a rather precise kind of scenario.
For the fermion DM case, there is only one decay possibility,
$\psi_{DM}\rightarrow \psi \gamma$ with $\psi_{DM}$ and $\psi$ necessarily carrying the same millicharge.
In the following, when establishing the list of operators 
that could lead to a sizable monochromatic photon signal, we will not specify the exact nature of the fermionic partner of the DM in these operators. It could be either a Dirac or a Majorana fermion, and it could be either a particle beyond the SM or a neutrino. 
The former option points towards a multi-component fermion DM scenario. 
Note that the results obtained below, in particular those of Fig.~\ref{fig:Kin_mix}, do not depend on how the abundancies of these components pile up to saturate the observed value of $\Omega_{DM} h^2=0.12$,  except those depending on the direct detection constraints on a millicharge.

Up to dimension six, there is only a very limited number of operators that can induce a $\psi_{DM}\rightarrow \psi \gamma$ decay from the millicharge of $\psi_{DM}$ and $\psi$. 
First of all, we only find a single dimension-five operator
\begin{equation}
 D_\mu  D_\nu  \bar{\psi}  \sigma^{\mu \nu} \psi_{DM} \,. \label{OFDpsisig}
\end{equation}
For this operator and all operators below, the addition of its hermitian conjugate is implicit.\footnote{Operators with covariant derivative(s) on the scalar field as  $\bar{\psi} \gamma_\mu \psi_{DM}  D^\mu \phi$
do not give any radiative two-body decays because this would require that the scalar field has both a vev and a millicharge, which would give a mass to the photon. Similarly, operators with a $\slashed{D} \psi$ or $D^2 \phi$ do not give any radiative decays as can be seen from the use of the equations of motion. Note also that operators with an additional $\gamma_5$ are redundant since both fermions in the operator are different fields (i.e.~it can always be reabsorbed in the definition of one of the fermion field).} The presence of the $\sigma_{\mu \nu }$ implies that this operator can be rewritten as a sum of operators where both covariant derivatives have been replaced by a sum of field strengths of the gauge boson to which the particle couples (each one multiplied by the corresponding gauge
coupling).\footnote{This basically means that such operator could be easily produced from one loop diagrams involving UV particles, in a way similar to the ones generating the  
usual $F^{\mu\nu}\bar{\psi}\sigma_{\mu\nu}\psi'$ dipole operators (as relevant for example for the $\mu\rightarrow e \gamma$ decay), with the difference that the photon would here be radiated by a millicharged particle instead of a charged lepton or charged gauge boson.}

As for the dimension-six operators we find only three possibilities, 
\begin{eqnarray}
D_\mu D_\nu \bar{\psi}  \sigma^{\mu \nu} \psi_{DM}  \phi\,, \label{O6FDpsisig}\\ 
\bar{\psi} \sigma^{\mu \nu} D_\mu D_\nu  \psi_{DM} \phi\,, \label{O6FDDMsig}\\
D_\mu \bar{\psi}  \sigma^{\mu \nu} D_\nu\psi_{DM}  \phi \,.\label{O6FD2sig}
\end{eqnarray}

Their structure are equivalent up to one operator that does not produce monochromatic photons but can give other decays (including two-body decays), hence different amounts of cosmic rays. As for the operator of Eq.~(\ref{OFDpsisig}), the covariant derivative of the operators of Eqs.~(\ref{O6FDpsisig}, \ref{O6FDDMsig}) 
can be traded for a sum of field strengths.
In summary, up to dimension six, we are left with four operator structures only, as given in Eq.~(\ref{OFDpsisig}) and Eqs.~(\ref{O6FDpsisig})-(\ref{O6FD2sig}).
At the two-body decay level, the scalar field in the last three operators can intervene only through its vev. 
For the quantum numbers of these fields there is in principle an infinity of possibilities  and we will see how, when considering the constraints on the various operators,  a simple global picture can emerge despite of this fact.

\section{Constraints on the various operators\label{Constraints}}

As for the non-millicharged operators in Ref.~\cite{Gustafsson::2013}, there are a priori essentially two main ways to constraint the operators and thus to possibly discriminate between them, from $\gamma$-ray line spectral features and from the associated continuum of cosmic rays produced. The fact that the lifetime of the DM particle must be longer than the age of the Universe provides an additional constraint which is relevant in special cases. 

By spectral features we mean the number of $\gamma$-ray lines produced, their relative energies and relative intensities. However, for the millicharged operators there is no way to get more than one $\gamma$-ray line from a unique given operator since in the final state, on top of the photon, one can only find the $\psi$ particle. 

To determine what are the possibilities to distinguish among the various operators from a cosmic ray multi-channel analysis, we will proceed as for the case of non-millicharged operators in Ref.~\cite{Gustafsson::2013}. 
The whole issue is that  for a given operator, due to gauge invariance, there always is a continuum flux of cosmic rays associated to the production of a $\gamma$-ray line, especially if its energy is larger than the $Z$ boson mass. 

The decay rate of the DM particle into photons is proportional to its millicharge squared. 
Therefore, to determine the upper bounds existing on the photon over cosmic rays ratios, we will need to know (in some cases) what are the bounds which hold on the millicharge of a metastable particle within the mass range we consider, $m_{DM}= \mathcal{O}(100-\hbox{few}~10^4)$ GeV.
In this mass range, there are no relevant accelerator constraints  
\cite{Redondo::2011,Fairbairn::2007,Davidson::2000, Prinz::1998}.  
There are nevertheless stringent constraints from cosmology as well as from direct detection data.
As well-known, in the usual $\Lambda$CDM model, which fits well both the CMB anisotropy and large scale structure data, it is assumed that there is no DM-baryon interaction other than gravitational. An additional DM-baryon interaction such as the one provided by a millicharge  modifies this picture by rendering DM effectively ``baryonic''. This affects the CMB power spectrum as well as the baryon acoustic oscillations, leading to the upper bound \cite{Dubovsky::2001,McDermott:2010pa,Dolgov::2013,Dvorkin::2013}
\begin{equation}
\frac{\sigma_0 }{M_{DM}}\leq 1.8\times 10^{-17} \text{cm}^2/\text{g}\,,
\label{CMB_bound}
\end{equation}
where $\bar{\sigma}=\sigma_0 v^n$ is the DM-baryon momentum-transfer cross-section and $v$ is the DM-baryon relative velocity. In our case, the relevant cross-section is the Rutherford one and $n=-4.$ Eq.~(\ref{CMB_bound}) translates into the following bound on the DM millicharge
\begin{equation}
Q^2_{DM} \leq 3.24 \times 10^{-12}  \alpha\,  (\frac{M_{DM}}{\text{GeV}}) \,,
\label{mc_bound}
\end{equation}
where $\alpha$ is the electromagnetic fine structure constant.

The direct detection bounds are much more model-dependent than the CMB ones. They crucially depend on the mass of the particle exchanged between the nucleon and the DM particle.
In the massless $U(1)'$ gauge boson case, the elastic nucleon-DM scattering is proportional to the inverse of the recoil energy squared, $d\sigma_N/dE_r \propto 1/E_r^2\sim 1/\hbox{KeV}^{2}$. This results in a huge enhancement of the scattering cross section with respect to the usual WIMP case where the cross section is typically suppressed by the inverse of the square of the GeV-TeV mass of the particle exchanged. For $m_{DM}\gtrsim $~few~GeV this results in upper bounds on the millicharge of order of $10^{-9}$-$10^{-10}$, see Fig.~9 of Ref.~\cite{Chu::2012} (where the $\kappa$ parameter is nothing but $\varepsilon \sqrt{\alpha'/\alpha}$ in our notation).
For the Stueckelberg case, these constraints remain also valid as long as the $Z'$ is lighter than $\sim10$ MeV, see for example Fig.1 of Ref.~\cite{Fornengo:2011sz}. Beyond these values, the upper bound on the millicharge scales as $1/m_{Z'}^2$, so that it quickly becomes weaker than the CMB bound (this occurs for $m_{Z'}\gtrsim 1$ GeV). In the following, when the bound on the millicharge is relevant to determine the upper bound on the $\gamma$-ray line, we use the CMB bound. We will see that this bound already excludes an observable $\gamma$-ray line for these cases (hence a fortiori when the direct detection bounds are stronger, this is even more excluded). Note also that all these direct detection constraints assume a standard local DM density, which might not apply depending on how large the DM particle millicharge is, because it can be shielded by galactic magnetic fields \cite{Chuzhoy:2008zy,McDermott:2010pa}. For instance, when $Q_{em}> 10^{-10}\cdot (m_{DM}/100\,\hbox{GeV})$, the depletion of the local DM relic density from magnetic shielding begins to be sizeable, therefore weakening the direct detection bounds.

The photon over cosmic rays ratio in both kinetic mixing and  Stueckelberg scenarios is given by 
\begin{equation}
\frac{n_\gamma}{n_{CR}}=\frac{Q^2_{DM}}{D} \,,
\label{ratio_DDpsidpsi}
\end{equation}
\begin{eqnarray}
D&=&  c^2_Z f_Z(M_{DM},M_Z) n_{CR/Z}+c^2_{Z'} f_{Z'}(M_{DM},M_Z')  n_{CR/Z'} \nonumber\\
&+& \frac{g^2}{4}c_W  f_W(M_{DM},M_W) \nonumber\\ 
&\times &(n_{CR/W^+}+n_{CR/W^-}+n_{CR/\psi^+}+n_{CR/\psi^-}). \nonumber
\end{eqnarray}
Here, the $n_{CR/P}$ ratios hold for the number of cosmic rays (of a given type and of a given energy) produced per particle P, and $f_{Z,Z',W}(M_{DM},M_{Z,Z',W})$ are functions of the $DM$ and $Z,Z',W$ masses. For $m_{DM}>>m_{Z,Z',W}$, they always are equal to unity except for the operator of Eq.~(\ref{O6FD2sig}), see below.
In these equations $Q_{DM}$ is the millicharge which, as said above, is equal to $Q_{DM} = \frac{- \epsilon g' Q'}{g_Y} e'$ and $Q_{DM}= \frac{- g Q' \tan(\phi) e'}{g_Y}$ in the kinetic mixing and Stueckelberg cases, respectively.
As for the coupling to the $Z$, $c_{Z}$, as Eqs.~(\ref{lagrangian_kin_mix})-(\ref{lagrangian_Stuck}) show, it takes the simple form
\begin{eqnarray}
c_Z &=& \frac{g T_3}{\cos(\theta_\epsilon)}+\frac{g' Q' \epsilon \sin(\theta_\epsilon)}{\sqrt{1-\epsilon^2}} \,,\label{cZkm}\\
c_Z &=&- g'Q' (s_\psi c_\phi + c_\psi s_\theta s_\phi)\nonumber\\
&&-g T_3 c_\theta  c_\psi \Big(1+t_\theta^2(1-\frac{t_\psi t_\phi}{ s_\theta})\Big) \,,\,\, \label{cZstu}
\end{eqnarray}
respectively (where $t_{\theta,\phi,\psi}$ indicates the tangent of $\theta, \phi, \psi$).
The coupling to the $Z'$, which applies only in the Stueckelberg case, is 
\begin{eqnarray}
c_{Z'}&=&- g'Q' (s_\psi s_\theta s_\phi - c_\psi c_\phi) \nonumber\\
&&-g T_3 s_\psi c_\theta \Big(1+t_\theta^2(1+\frac{t_\phi}{ s_\theta t_\psi})\Big).
\end{eqnarray}
Finally, the coupling to the $W$, $c_W$, can take very different values as a function of the multiplets considered in the various operators (and the associated Clebsch-Gordan coefficients). In practice, we will consider the cases which, among all possible multiplet configurations up to $SU(2)_L$ quintuplets, minimize $c_W$, hence maximize the $n_{\gamma/CR}$ ratio.

Before coming to the constraints which hold for the various operators, note that in the following we will not consider in many details the amount of cosmic rays the $Z'$ could give. The limits on the intensity of the $\gamma$-ray line we will give below hold for the case where the $Z'$ does not give any cosmic rays. This can be the case for example if the $Z'$ decays essentially to $\psi \bar{\psi}$ (if $M_{Z'}\geq 2 M_{\psi}$). This situation gives the maximum upper bound that could be reached. When neglecting the contribution of the $Z'$ to the cosmic rays, both Stueckelberg and kinetic mixing scenarios give rise to the same  Fig.~\ref{fig:Kin_mix}  bounds at the lowest order in $\phi$ and $\epsilon$ respectively. At the end of Sec. \ref{Constraints}, we will discuss how our results might be affected if, instead, the $Z'$ mainly decays into SM particles.

\subsection{Constraints on the $D_\mu D_\nu \bar{\psi} \sigma^{\mu \nu} \psi_{DM } $ operator}
For the unique dimension-five operator, the quantum numbers of $\psi_{DM}$ and $\psi$ are necessarily the same, and in particular  $T^{DM}_3=T^{\psi}_3$.
The simple crucial remark to be done at this stage, not only valid for this operator but for all operators, is that unless the fields to which  the covariant derivative applies is a singlet of both $SU(2)_L$ and $U(1)_Y$, there will always be a two-body decay production of a $Z$ and/or a $W$ in a way which is not suppressed by the value of the millicharge.
As a result in this case, the production of cosmic rays is boosted, by the inverse of the millicharge squared, with respect to the production of the $\gamma$-ray. 
Z bosons are produced in this way as soon as the $T_3$ of the particle to which  the covariant derivative applies is different from zero, see Eqs.~(\ref{cZkm}) and (\ref{cZstu}).
As for the $c_W$ coupling, for the dimension-five operator one can write it as $c_W=a+b$. For an $SU(2)_L$ multiplet of dimension $n= 2 \lambda +1$, $a=0$ if $Y_{DM}= 2 \lambda$ and $a=1$ in all other cases. Similarly, $b=0$ if $Y_{DM}=- 2 \lambda$ and $b=1$ in all other cases.
As a result, both $a$ and $b$, hence $c_W$, can be zero only if $\lambda=0$, that is to say if one considers a SM singlet.

From this discussion there are three general cases
\begin{itemize}
\item \underline{$\psi_{DM}$ and hence $\psi$ are SM singlets}. In this case there is no $W$ production and the $Z$ production involves two powers of the millicharge, as for the $\gamma$ production. As a result we get an unsuppressed ratio
\begin{equation}
\frac{n_\gamma}{n_{CR}} \simeq \frac{1}{\tan^2 \theta_W n_{CR/Z}}\,,
\label{ratio_A}
\end{equation}
at the lowest order in $\epsilon$ for the kinetic mixing case and in $\phi$ for the Stueckelberg case.  In the latter case, this is a good approximation if $\sin \phi << \sin \psi$, which is what is expected if there is a big splitting between the SM gauge bosons and the $Z'$ masses. 

\item \underline{$T_3^{DM}=T_3^{\psi}=0$ with $\psi_{DM}$, $\psi$ non-singlets}.  In this case, there is production of $W$ bosons and  $a+b=2$ in Eq. (\ref{ratio_DDpsidpsi}), meaning that
\begin{equation}
\frac{n_\gamma}{n_{CR}}=\frac{ Q^2_{DM}}{\frac{g^2}{2} (n_{CR/W^+}+n_{CR/W^-}) } \,,
\label{ratio_B}
\end{equation}
at the lowest order in $\epsilon $ or $\phi$.
Here, in order to obtain a conservative model-independent upper bound, we made the hypothesis  that the charged $\psi$ components produced in two-body decays together with $W$ bosons do not yield an important contribution to cosmic rays production. 
\item \underline{$T_3^{DM}=T_3^{\psi}\neq0$}. In this case we have both unsuppressed production of $Z$ and $W$. Here we will consider only the case with $a+b=1$ and $T^{DM}_3=1/2$, as it is the one which maximizes the $n_\gamma/n_{CR}$ ratio of Eq.~(\ref{ratio_DDpsidpsi}). It gives
\begin{equation}
\frac{n_\gamma}{n_{CR}}=\frac{ Q^2_{DM}}{c_Z^2 n_{CR/Z}+\frac{g^2}{4} (n_{CR/W^+}+n_{CR/W^-}) } \,,
\label{ratio_C}
\end{equation}
with $c^2_Z \simeq  \frac{g^2}{4 \cos^2 \theta_W}$ at the lowest order in $\epsilon $ or $\phi$.
\end{itemize}

\subsection{Constraints on the $D_\mu D_\nu \bar{\psi} \sigma^{\mu \nu} \psi_{DM } \phi $ operator}
In this case, the relevant quantum numbers are those of $\psi$. The maximal ratios as a function of the value of  $T_3^\psi$ follow the same pattern as for the previous operator. In the same way, one has three cases
\begin{itemize}
\item \underline{$\psi$ is a SM singlet}: prediction (\ref{ratio_A}).
\item \underline{$T_3^\psi=0$ with $\psi$ non- singlet}: prediction (\ref{ratio_B}).
\item \underline{$T_3^\psi\neq 0$}: prediction  (\ref{ratio_C}). \end{itemize}

\subsection{Constraints on the $ \bar{\psi} \sigma^{\mu \nu} D_\mu D_\nu \psi_{DM } \phi $ operator}
Now, the relevant quantum numbers are those of the DM particle.   Here, the three cases are:
\begin{itemize}
\item \underline{$\psi_{DM}$ is a SM singlet}: prediction of Eq. (\ref{ratio_A}). 
\item \underline{$T^{DM}_3=0$ with $\psi_{DM}$ non- singlet}:  the ratio is maximized with $ c_W =1/2$, which gives\footnote{This minimum value of $c_W$ is obtained for a situation with $\psi_{DM}$ a triplet, and $\psi$ and $\phi$ quintuplets. Any combination of smaller multiplets gives a bigger value of $c_W$. For example, taking $\psi_{DM}$ as a hyperchargeless triplet, and both $\psi$ and $\phi$  as doublets of hypercharge 1,  gives $c_W=2$.} 
\begin{equation}
\frac{n_\gamma}{n_{CR}}=\frac{ Q^2_{DM}}{\frac{g^2}{8} (n_{CR/W^+}+n_{CR/W^-}) }
\label{ratio_D}\,.
\end{equation}
\item \underline{$T_3^\psi\neq 0$}: the ratio is maximized with $T_3^\psi =1/2$ and $c_W =1/4$, which gives
\begin{equation}
\frac{n_\gamma}{n_{CR}}=\frac{ Q^2_{DM}}{c_Z^2 n_{CR/Z}+\frac{g^2}{16} (n_{CR/W^+}+n_{CR/W^-})}. 
\label{ratio_E}
\end{equation}
\end{itemize}

\subsection{Constraints on the $D_\mu  \bar{\psi} \sigma^{\mu \nu}  D_\nu \psi_{DM } \phi $ operator}

The phenomenology of this operator is more involved than the one of the operators above because it depends in a complicated way on the couplings of both $\psi$ and $\psi_{DM}$ to the various bosons. Nonetheless, maximizing the $n_\gamma/n_{CR}$ ratio requires $\psi$ and $\psi_{DM}$ to have the same quantum numbers, which greatly reduces the complexity of the $n_\gamma/n_{CR}$ ratio.

In the case where both $\psi_{DM}$ and $\psi$ are SM singlets, this ratio is given by
\begin{equation}
\frac{n_\gamma}{n_{CR}} \simeq \frac{1}{\tan^2\theta_W(1-(\frac{M_Z}{M_{DM}})^2)^2 (1+\frac{1}{2}(\frac{M_Z}{M_{DM}})^2)~ n_{CR/Z}}\,,
\label{ratio_F}
\end{equation}
at the lowest order in $\epsilon$ or $ \phi. $ 

In the case where $\psi_{DM}$ and $\psi$ are not SM singlets, the predictions are quite lengthy and we only give them for the cases where $T_3=0$ and $T_3=1/2$ in the Appendix. Unlike  all previous cases where the dependence on $m_W$ is negligible for $m_{DM}>>m_W$, here, from the longitudinal $W$ contribution, there are terms in $m_{DM}/m_W$ which imply a power law dependence on $m_{DM}$, see Appendix. 
This is associated to the fact that, unlike the other operators, this one is not equivalent to  a single operator involving  a field strength. 
This will give rise to a different behaviour of the constraints below, as the bounds obtained for the $D_\mu  \bar{\psi} \sigma^{\mu \nu}  D_\nu \psi_{DM } \phi $ operator get stronger when $m_{DM}$ increases.

\subsection{Results}
\label{Results}

\begin{figure}[h]
\center{\includegraphics[width=1 \columnwidth]{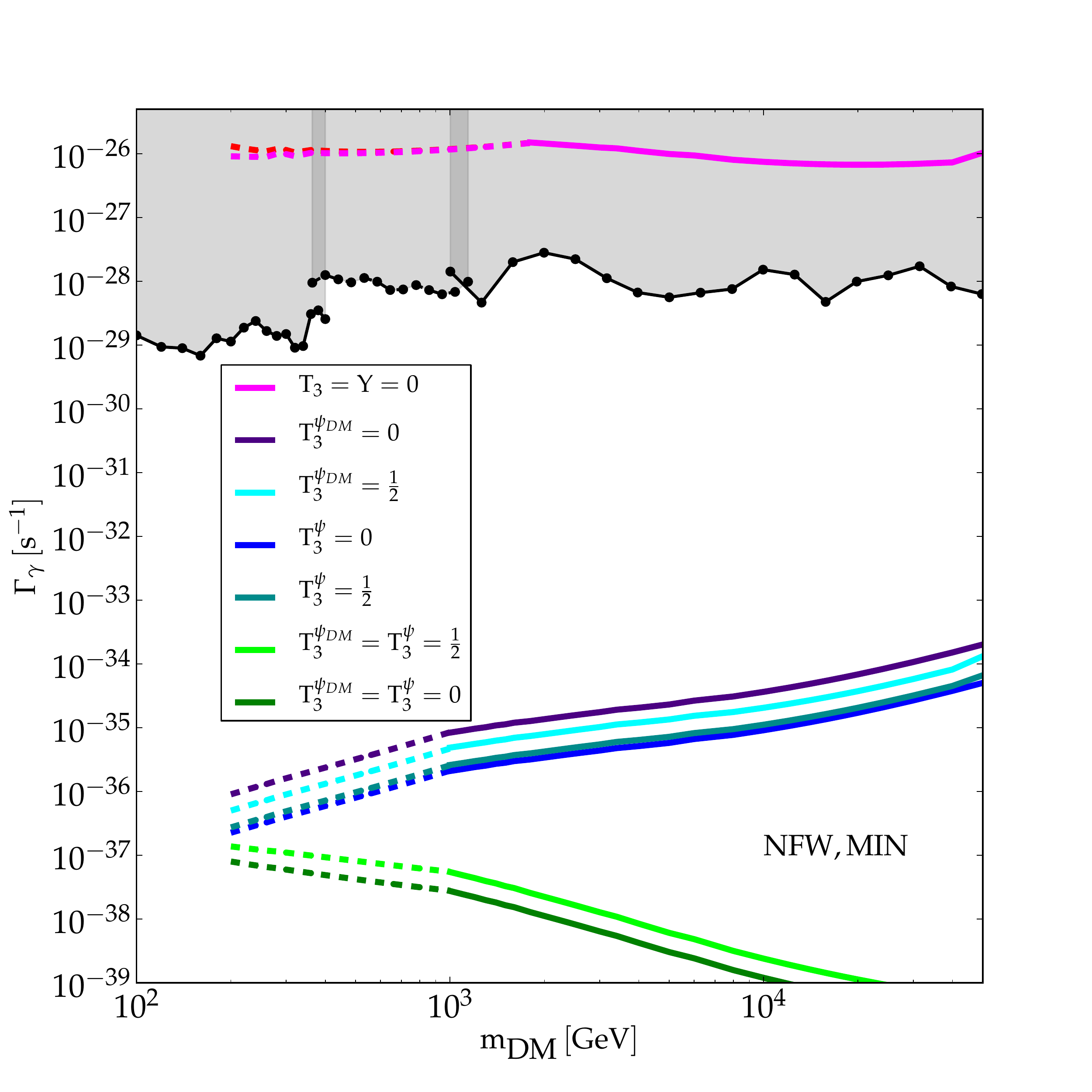}}
\caption{\footnotesize{Upper bounds on the decay rate into monochromatic photons from the predicted $n_{\gamma/CR}$. Dashed curves represent limits from PAMELA measurements on the $\bar{p}$ flux, and continuous curves are constraints derived from measurements of diffuse photon background of Fermi-LAT. Grey areas are excluded by direct searches from Fermi-LAT and HESS experiments. We considered a NFW profile \cite{Cirelli::2012} for the DM density and used the MIN propagation model to compute conservative $\bar{p}$ constraints \cite{Cirelli::2012}. Using instead the MAX profile, the bounds for the $\bar{p}$ would go down by approximately one order of magnitude. The $T_3=Y=0$ labelled curve is the upper bound for almost all operators when the DM is a SM singlet, Eq.~(\ref{ratio_A}). The only exception is the $D_\mu  \bar{\psi} \sigma^{\mu \nu}  D_\nu \psi_{DM } \phi $ operator with a singlet DM, whose bound is almost identical except at DM masses around 200 GeV, where it is given by the red curve. $T_3^\psi$ labelled curves are limits on the $D_\mu D_\nu \bar{\psi} \sigma^{\mu \nu} \psi_{DM } \phi$ and $D_\mu D_\nu \bar{\psi} \sigma^{\mu \nu} \psi_{DM } $ operators, whereas $T_3^{\psi_{DM}}$ labelled curves are limits on the  $ \bar{\psi} \sigma^{\mu \nu} D_\mu D_\nu \psi_{DM } \phi $ operator. The two remaining curves, labelled  $T_3^{\psi_{DM}}=T_3^{\psi_{DM}}$, correspond to the operator $D_\mu  \bar{\psi} \sigma^{\mu \nu}  D_\nu \psi_{DM } \phi $. These constraints hold for the kinetic mixing as well as for the Stueckelberg frameworks at first order in $\epsilon$ and $\phi$. They also apply in the ``just-so'' millicharge scenario.}}\label{fig:Kin_mix}
\end{figure}

Fig.~\ref{fig:Kin_mix} shows the constraints obtained on the photon over cosmic rays ratios for the different operators involving a millicharged DM. For low masses, up to $\simeq 1$ TeV, the most stringent bounds are provided by PAMELA measurements of cosmic antiprotons \cite{PAMELA::2010} , whereas for higher masses, the relevant constraints come from measurements of the diffuse gamma background from Fermi-LAT \cite{Fermi::2010a} and HESS \cite{HESS::2013}. The methodology used to obtain these constraints is the same as in Ref. \cite{Gustafsson::2013}. By comparing the constraints from cosmic rays on $n_{\gamma}/n_{CR}$ and the limits from direct searches for photon line spectral features, this plot shows which operator are compatible with a possible near future observation of a $\gamma$-ray line.

A clear pattern emerges from these results.  Except for the case in which the DM is a singlet of the SM gauge groups, none of the effective operators associated to a millicharged DM  taken individually would be able to produce a  $\gamma$-ray line strong enough to meet the current experimental sensitivity without overproducing antiprotons and diffuse photons.  Actually, in all the cases in which the DM is not a SM singlet, the observation of a  spectral photon line would be associated to an excess of cosmic rays from five to ten orders of magnitude higher than the present experimental bounds from PAMELA or Fermi-LAT and HESS. This result is interesting because it singles out a unique possibility for the SM quantum numbers of the DM: a SM singlet. Unfortunately, it does not allow to discriminate, neither among  the mechanisms responsible for the millicharge of the DM, nor  among the various operators. When the DM is a singlet, all operators lead to the prediction 
of Eq.~(\ref{ratio_A}), $n_{\gamma/CR} \simeq 1/(\tan^2\theta_W n_{CR/Z} )$,  except Op.(\ref{ratio_F}) which differs at DM masses close to $M_Z$, but only very slightly.  As Fig.~1 shows, this prediction gives the maximum $\gamma$-ray line intensity allowed by cosmic rays constraints. 
This absolute bound turns out to be the same as in Ref.~\cite{Gustafsson::2013} for the neutral DM scenarios.

As for a non-SM-singlet DM, in Fig.~1 we have only considered  the quantum numbers that maximize $n_{\gamma}/n_{CR}$. The maximum ratios turn out to be identical for operators $D_\mu D_\nu \bar{\psi} \sigma^{\mu \nu} \psi_{DM } $ and $D_\mu D_\nu \bar{\psi} \sigma^{\mu \nu} \psi_{DM } \phi$. They differ by less than one order of magnitude for $ \bar{\psi} \sigma^{\mu \nu} D_\mu D_\nu \psi_{DM } \phi $, which is within the uncertainty coming from the propagation models.  The only operator providing very different bounds for non-singlet DM is the operator $D_\mu  \bar{\psi} \sigma^{\mu \nu}  D_\nu \psi_{DM } \phi $, due to the dependence of the $n_{\gamma}/n_{CR}$ ratios on the DM mass. Not only are these bounds exhibiting a different behaviour for increasing $m_{DM}$, but they also differ by more than two orders of magnitude from all the previous bounds for $m_{DM} \gtrsim 2$ TeV.  Therefore, if a line were to be detected with a sensitivity of direct searches for monochromatic photons improved by several orders of magnitude, it would in principle be possible to discriminate this particular operator from the other ones where the DM is not a SM singlet. But, in practice, this does not appear at all to be a realistic option because this basically means a $\gamma$-ray line with intensity smaller than the intensity of the photon continuum observed. Putting together the results obtained for a millicharged DM and those derived for a neutral DM \cite{Gustafsson::2013}, we find that if a line were detected at the current experimental sensitivity without any excess of cosmic rays, it would not be possible to discriminate the millicharged SM-singlet scenario from the neutral DM case giving the same Eq. \ref{ratio_A}  ratio (prediction ``A'' in \cite{Gustafsson::2013}). However, if on the contrary a strong line were to be detected with a sizeable associated cosmic rays excess, only the more suppressed $\gamma$-ray predictions ``B''-``E'' in Ref. \cite{Gustafsson::2013} for neutral DM  could explain it.

As said above, all the bounds obtained have been computed under the hypothesis that the Stueckelberg $Z'$ that could be present in two-body decay final states (if kinematically allowed) does not produce any cosmic rays by decaying subsequently.  If this does not hold, the operators will give rise to more suppressed bounds.
To estimate how important this contribution could be, we consider as an example, with $g'Q'=1$, a $Z'$ boson which decays mainly to $b \bar{b}$, a channel which is known to produce many cosmic rays.
When the DM is not a singlet, we find bounds that are stronger by approximately two orders of magnitude. 
This stems from the fact that, in this case, a single $Z'$ produces a comparable amount of cosmic rays with respect to  the SM gauge bosons, but, with $g' Q'=1$,  its coupling  to the DM particle is stronger than those of the  $Z$ or $ W$ bosons. If, instead, the DM particle is a SM singlet,  the $n_\gamma/ n_{CR}$ ratio does not depend anymore on the value of $g'Q'$, but the coupling to the photon is suppressed by $\sin\phi$.  Instead of having a fixed bound on the ratio as given by Eq.~(\ref{ratio_A}),
 the following limit is obtained:
\begin{equation}
\frac{n_\gamma}{n_{CR}} \lesssim \frac{\cos^2\theta \sin^2\phi}{n_{CR/Z'}}\,.
\end{equation} 
The $\phi$ angle is constrained by the measurement of the $Z$ width from LEP,  $\sin\phi \leq 0.04$ \cite{Feldman::2007}. This decreases the bound of Eq.~(\ref{ratio_A}) by 
three to four orders of magnitude. Interestingly,  the observation of a $\gamma$-ray line with intensity of order the present experimental sensitivity would therefore probe this possibility, see Fig.~1.

\begin{figure}[h]
\center{\includegraphics[width=1 \columnwidth]{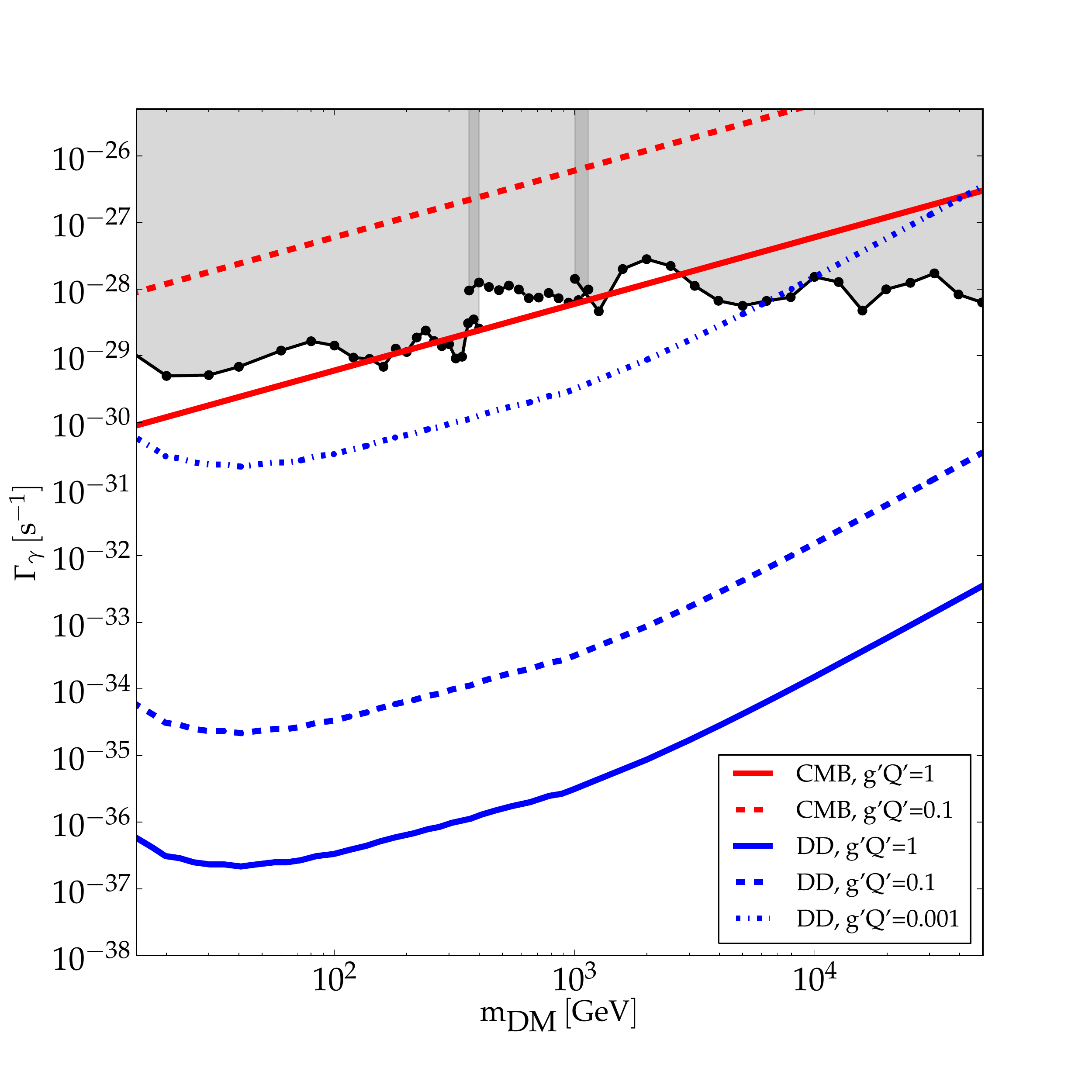}}
\caption{\footnotesize{For the kinetic mixing case, bounds on the intensity of the $\gamma$-ray line as a function of $m_{DM}$ assuming various values of $g'Q'$, imposing $\Gamma^{-1}(\psi_{DM}\rightarrow \psi \gamma')>\tau_U$, and considering the CMB bounds of Eq.~(\ref{mc_bound}) (red curves) and direct detection bounds (DD, blue curves) on the millicharge. The direct detection bounds we used are those from Xenon100, Fig.~9 of Ref.~\cite{Chu::2012}.}}\label{fig:Kaaaaa}
\end{figure}

As mentioned above for the massless kinetic mixing case, the direct detection constraints on the millicharge are stronger than the ``universal'' CMB constraints we have used for Fig.~\ref{fig:Kin_mix}. As a result, in this case, an observable $\gamma$-ray line for the non-SM singlet scenario is even less of an option, not only because this would give even more cosmic rays, but also because it would give a DM lifetime smaller than the age of the Universe.
Concerning possible rapid decays,  note also that since the radiative decays are suppressed in all cases (singlet case included) by a factor of the millicharge squared, if there exist other (non-radiative) operators induced at the scale $\Lambda$ which destabilize the DM particle in a way which is not suppressed by this factor, they could easily induce decays much faster than the radiative ones. These decays could induce cosmic ray fluxes above the ones observed or even give a DM lifetime shorter than the age of the Universe. 
The scenario is therefore viable if there is no such operator or if the associated decays do exist, but with a lifetime larger than the age of the Universe and without an excessive associated production of cosmic rays.

So far, we have discussed the constraints on the intensity of a $\gamma$-ray line which hold from the bounds on the millicharge itself, as this is the parameter entering in the various rates.
However, if one has some prejudice about the values of the parameters responsible for the millicharge,  there are cases where stronger bounds on the $\gamma$-ray line intensity apply. This is relevant in particular for the kinetic mixing case if, as for any gauge charge in the SM, one considers a value of the ``dark charge'' $g'Q'$ not far from unity. The issue here is that in the massless hidden gauge boson case, the DM particle does not  only decay to $\psi + \gamma$ but it also does to $\psi +\gamma'$. As the $\gamma'$ does not produce any cosmic rays, the latter decay is not relevant for Fig.~1. However, it is naturally faster than the former decay by a factor of $\epsilon^{-2}$. One has therefore to make sure that the resulting lifetime is not shorter than the age of the Universe. Imposing that $\Gamma^{-1}(\psi_{DM}\rightarrow \psi \gamma')>\tau_U$, and imposing that the millicharge satisfies the CMB bound of Eq.~(\ref{mc_bound}),  we show in Fig.~2 the upper limits which hold on the intensity of the $\gamma$-ray line for various values of the dark charge $g'Q'$. Clearly, for large values of $g'Q'$ one gets competitive bounds with respect to  those derived from $\gamma$-ray line direct searches, whereas smaller values give irrelevant bounds. Fig.~2 also shows the constraints we get assuming the direct detection bounds mentioned above (i.e.~Fig.~9 of Ref.~\cite{Chu::2012}, disregarding possible weakening of these bounds from magnetic shielding). These constraints are quite stringent, leading to unobservable $\gamma$-ray lines, unless the dark charge is small enough. In fact, $g'Q' \lesssim 10^{-3} (m_{DM}/\text{TeV})$ is necessary in order to get $\Gamma (\psi_{DM}\rightarrow \psi \gamma) \gtrsim 10^{-30} \hbox{sec}^{-1}$.  These constraints are obtained assuming that the relic density of $\psi_{DM}$ is the observed one.  For smaller relic density values one gets weaker constraints. In the Stueckelberg case, all these considerations become obviously irrelevant as soon as the $Z'$ gauge boson mass is above $m_{DM}$ (or above $\sim$~GeV for what concerns the direct detection bounds).

\section{What about the $\gamma+\nu$ option?} 

From the above results it is clear that, since the neutrino is not a SM singlet, it is not an option within the $m_{DM}={\it O}(100-\hbox{few}\,10^4)$~GeV range considered above. Moreover in this case, since the millicharge of both fermions in the various operators must be equal, 
and since the millicharge of the neutrinos is extremely well bounded, the millicharge of the DM particle is also extremely well constrained.
The most stringent constraint applies on the electronic neutrino. Assuming charge conservation in $\beta$ decay, and using the experimental results from \cite{Marinelli::1984} and \cite{Baumann::1988}: $q_{p^+}+q_{e^-}= (0.8\pm 0.8) 10^{-21} e$  and $q_{n^0}= (-0.4\pm 1.1)10^{-21} e$,  the constraint $q_{\nu_e} \lesssim10^{-21} e$ is obtained.  Independent, less stringent upper bounds also hold from neutrino magnetic dipole moment searches, see \textit{e.g.} \cite{Gninenko::2007}.  In the case of $\nu_\mu$ and $\nu_\tau$, the most stringent constraints come from stellar evolution \cite{Raffelt::1999}. If neutrinos acquire a millicharge, their electromagnetic interactions would provoke extra energy losses in the core of red giants. This would delay the time of helium ignition, and as a consequence, the core of red giants would be heavier than in the standard case when helium lights up. But the mass of the red giant core at helium ignition is constrained by measurements from globular clusters. These constraints turn into the following  bound on the charge of neutrinos: $q_\nu \lesssim 2 \times 10^{-14} e$. This bound holds as long as $m_\nu \lesssim 5$ KeV, implying that it  applies to all flavours.  To get a lifetime allowing for an observable $\gamma$-ray line, the huge suppression due to this millicharge could in principle be compensated by considering smaller $\Lambda$ scales, taking typically $\Lambda_{GUT} \simeq 10^{15}$ GeV for dimension-five operators, and $\Lambda \simeq 10^{8}$ GeV for dimension-six operators.
However, since the $Z$ emission rate is enhanced with respect to the $\gamma$ emission by a factor of $1/Q^2_{\nu}$, this would clearly imply a decay rate to $Z+\nu$ leading to a lifetime much shorter than the lifetime of the Universe. Similarly this would have given 
a huge amount of cosmic rays.\footnote{For the neutrino case we should take the bounds obtained in the case with $T^\psi_3=1/2$ in Fig. \ref{fig:Kin_mix} and rescale them by a factor of $Q^2_{DM}/Q^2_\nu$ where $Q^2_{DM}$ refers to the bound of Eq.~(\ref{mc_bound}).} Therefore, due to several reasons, a line observed at the present sensitivities with energies above the $Z$ mass could not be attributed at all to a millicharged DM decaying in neutrino and photon through one of the operators under study.

More generally, one could ask whether this possibility is also excluded for lower DM masses. Here, there are a priori two directly connected constraints which must be fulfilled, giving an upper and a lower bound on $m_{DM}$. 

The upper bound comes from the fact that, even if for $m_{DM}< m_Z$ the $Z$ cannot be produced on its mass shell, it can be produced off-shell and subsequently decay to a pair of fermions. This could result in a lifetime shorter than the age of the Universe and/or to too many cosmic rays. For instance, supposing that a $\gamma$ line, to be observable, must result from a two-body decay typically giving a lifetime
$\tau_\gamma\sim 10^{26-30}$~sec, it is easy to see that the three-body decay lifetime (from $DM\rightarrow 3 \nu$) is shorter than the age of the Universe, unless
 \begin{equation}
 m_{DM}< 35~\hbox{MeV}\cdot \Big(\frac{10^{28}\,\hbox{sec}}{\tau_\gamma}\Big)^{1/4}. 
 \end{equation}
The limit coming from the single neutrino channel gives therefore a strong enough limit on $m_{DM}$ to render irrelevant the limits one could get from all other possible $Z$ decay channel, except from the $\psi_{DM}\rightarrow \nu e^+ e^-$ channel.
It is easy to see that the latter channel gives up to a very good approximation the  stronger constraint, 
\begin{equation}
m_{DM}< 2 \,m_e\,,
\end{equation}
which holds in order to avoid overproduction of galactic center $511$~KeV photon from overproduction of positrons. The limit on the lifetime of a DM particle with mass below $\sim 35$~MeV is given by $\Gamma(\psi_{DM}\rightarrow \nu e^+e^-)< 10^{-26}\,s^{-1}\cdot(m_{DM}/\hbox{MeV})$ \cite{Bell:2010fk,Frigerio:2011in}.

As for the lower bound, it comes from the fact that, if one decreases too much $m_{DM}$, one gets a decay into a photon and a neutrino which is too slow to account for any observable photon line that could be detected in the future. For instance, let us consider the X-ray line recently reported with energy $\sim 3.5$~KeV and flux $F\simeq 10^{-6} \text{cm}^{-2} \text{s}^{-1}$ \cite{Bulbul:2014sua,Boyarsky:2014jta}. Assuming a standard DM density along the line of sight, such a line, if better confirmed experimentally, could be understood from a DM decay into a photon  and a neutrino, if the lifetime is  $10^{28}-10^{29}$ s \cite{Bulbul:2014sua,Boyarsky:2014jta}.
It is interesting to stress that such kind of lines could in principle be accounted for by any model leading to one of the fermion radiative operator reported in Ref.~\cite{Gustafsson::2013} (for the non-millicharged case) or by any of the millicharged operators considered here, provided the operator can match the constraint that $\psi$ must be a lepton doublet and $\psi_{DM}$ must be a singlet (as its mass must lie in the KeV range).\footnote{The list of operators given in Ref.~\cite{Gustafsson::2013}, given for a DM candidate above the $Z$ mass, also holds for lower mass.
To explain this recently reported line, one would not need necessarily to assume a fermion DM candidate. The scalar or vector operators given in this reference, or in section \ref{scalarvector} below, could also account for it, provided there exists another lighter scalar or vector particle to accompany the photon in the decay final state. To distinguish among these operators appears to be hopeless, given the fact that the associated neutrino flux is basically unobservable at these energies. For various other possible explanations of this line, see Refs.~\cite{Bulbul:2014sua,Boyarsky:2014jta,Ishida:2014dlp,Finkbeiner:2014sja,Higaki:2014zua,Jaeckel:2014qea}.}
For the millicharge option this latter requirement excludes the dimension-five operator but not the three dimension-six operators of Eqs.~(\ref{O6FDpsisig})-(\ref{O6FD2sig}) with $\phi$ the scalar SM doublet. 
For these operators one has nevertheless to check that for such low masses, and given the stringent constraints on the millicharge of neutrinos, one can get a lifetime of order the one needed.
The radiative lifetime one gets for any of these three operators is the same
\begin{eqnarray}
\tau_{\psi_{DM}\rightarrow \gamma\nu}&=& \frac{256  \pi \Lambda^4 }{Q_\nu^2 m_{DM}^3 v^2}\\
&=&  10^{29} \text{s} \Big(\frac{7\,\hbox{KeV}}{m_{DM}}\Big)^3 \Big(\frac{2 \cdot 10^{-14}}{Q_\nu}\Big)^2 \Big(\frac{\Lambda}{\hbox{600 GeV}}\Big)^4
\end{eqnarray}
which, for the parameter values indicated, is about the one needed. Given the uncertainties on the experimental flux needed, on the DM lifetime needed, and on the bounds on the neutrino millicharge from red giants, at the effective theory approach level one concludes that a small millicharge for the neutrino could be at the origin of this $\gamma$-ray line, or more generally of observable KeV-MeV low energy lines, provided there is new physics around the corner at colliders and provided that the millicharge of the $\nu_\mu$ or $\nu_\tau$ is close to its upper bound. In other words, despite of the very stringent bounds which exist on them, neutrino millicharges could consequently have an observable effect in the form of a X-ray line.  Note also that within the KeV-MeV mass range discussed here, for $m_{DM}\geq 50$ KeV, the generation of the relic density for a fermionic SM singlet DM can nevertheless be challenging \cite{Boyarski::2013}.

 Finally, remark that in the kinetic mixing scenario, imposing as above that $\Gamma^{-1}(\psi_DM\rightarrow \nu \gamma')>\tau_U$, gives the constraint $\varepsilon^2 > 5 \cdot 10^{-11} (10^{28} \,\hbox{sec}/\tau_\gamma)$, which means $g'Q'< 3 \cdot 10^{-9} (Q_\nu /2\cdot 10^{-14})$.

\section{Scalar DM and Vector DM\label{scalarvector}}

After studying  the possibility for a millicharged DM of the fermionic type to emit an observable $\gamma$-ray line through its decay,  we now  turn to  the scalar and vector DM cases.
If the DM particle is of the scalar or vector type, there also exist operators that could a priori  lead to such a line. For what concerns the emission of cosmic rays, the phenomenology of these operators turns out to be similar to the fermion case. Therefore, in the following, we will limit ourselves to the determination of the operators and to a few additional general comments which slightly distinguish these scenarios from the fermion case.

Due to angular momentum conservation considerations, a scalar DM particle cannot decay to a scalar and a massless gauge boson. Therefore, it can only decay to a photon and another massive spin-one particle, which necessarily carries the same millicharge. As a result, the scalar case is similar to the vector case. The operators that can allow a decay of a scalar DM particle into a vector particle and a photon, could also hold for a vector DM particle decaying into a scalar and a photon.
In principle, a vector DM particle could nevertheless also decay in a different way, into a photon and another vector particle. Note that a necessary condition for the vector DM to acquire a millicharge is to be a complex field, therefore associated to a non-abelian gauge group.

For the scalar-vector-photon case we found only one dimension-five operator
\begin{equation}
 F^A_{\mu \nu} F^{A \mu \nu}  \phi\,,  \label{OSV5}
\end{equation}
and two dimension-six operators
\begin{eqnarray}
 F^A_{\mu \nu} F^{ A \mu \nu}  \phi \phi' \,, \label{OSV6FF}\\
 F^A_{\mu \nu}  D^\mu \phi D^\nu \phi' \,.\label{OSV6FDD}
\end{eqnarray}
As said above, in these operators the DM particle can be either one of the scalar particle or a vector particle, the latter in one of the covariant derivative of Eq.~(\ref{OSV6FDD}) or in one of the non-abelian hidden sector $F^A_{\mu\nu}$ field strengths of Eqs~(\ref{OSV5})-(\ref{OSV6FF}). The photon can show up from one of the $F^A_{\mu\nu}$ field strengths, through gauge boson mixing. This occurs for instance if, on top of a kinetic mixing between the hypercharge gauge boson and a $U(1)'$ gauge boson, there is, through symmetry breaking, a mixing between this $U(1)'$ gauge boson and gauge boson(s) of the new non-abelian symmetry (whose field strength is $F^A_{\mu\nu}$). 
In this way the $F^A_{\mu\nu}$ field strength provides a photon field proportionally to the millicharge of the complex gauge boson in this  field strength.\footnote{Note that explicit realizations of such a possibility are nevertheless rather involved. For instance for the dimension-five operator of Eq.~(\ref{OSV5}) and a $SU(2)$ $F^A_{\mu\nu}$ field strengths, both field strengths must be taken in their quintuplet combination, which means that the scalar field is a quintuplet.}
Note that no operator containing only covariant derivatives remains. This can be shown using equations of motion and rotating away non-canonical kinetic terms. Operator (\ref{OSV6FDD}) is equivalent to Op. (\ref{OSV6FF}) up to operators that do not produce monochromatic photons. 

For the two operators of Eqs.~(\ref{OSV5})-(\ref{OSV6FF}), the emission of a $Z$ is always suppressed by the millicharge squared, even if the scalars are non-SM singlet. This stems from the fact that the $Z$ as the $\gamma$ can come only from the field strength in these operators, not from a covariant derivative as for the fermion operators. This means that one gets the Eq.~(\ref{ratio_A}) prediction even if the scalars are not SM singlets (up to corrections in $m^2_Z/m^2_{Z'}$ for $m^2_{Z'}> m^2_Z$). Of course, as for the fermion case, one can saturate the ratio of Eq.~(\ref{ratio_A}) only if there is no cosmic ray production from $Z'$ decay (in the Stueckelberg case). The latter could arise from the decay where the photon is replaced by a $Z'$ if this decay (which is not suppressed by the value of the millicharge squared) is not kinematically forbidden and if the $Z'$ subsequently decays into SM particles.
As for the operator of Eq.~(\ref{OSV6FDD}), it can lead to a non-millicharge suppressed production of $Z$ and $W$ from the covariant derivatives, if the scalars are non-SM singlets. Its phenomenology is therefore very similar to the one of the fermion operators.

Finally, about the possibility that a vector DM particle could decay into a photon and another vector particle, one could think about operators with only $F^A_{\mu\nu}$ field strengths, for instance $F^A_{\mu\nu}F^{A\mu\rho} F^{A\nu}_\rho$, through a similar mechanism where a $F^A_{\mu\nu}$ could give a photon through gauge boson mixing. We did not find  any simple realization of such a possibility.

\section{Summary}

In summary, there are very few ways of probing the DM hypothesis that can really be considered in a systematic and model-independent way.  However, for the decay of an absolutely neutral DM particle into a $\gamma$-ray line, this turns out to be feasible \cite{Gustafsson::2013}. This stems from the facts that, on the one hand, the use of an effective theory is fully justified, slow enough decay can naturally be  explained from a much higher scale physics, and, on the other hand, it turns out that there are very few 
operator structures of this kind. Ref. \cite{Gustafsson::2013}  considered the usual scenario where the DM particle is absolutely neutral so that the photon appears in the operator through a field strength (i.e. typically from a charged particle in a loop). Here we show that, for the same reasons, such a study can also  be systematically carried out  in the less considered scenario where the DM is millicharged, having therefore a tree-level coupling to the photon through a covariant derivative, either from an  ad-hoc millicharge, or through mixing of the $U(1)_Y$ gauge boson with another $U(1)'$ gauge boson.

To the emission of a $\gamma$-ray line from such operators is associated the emission of a continuum of cosmic rays.
The monochromatic photon to cosmic ray flux ratio is determined by the SM quantum numbers of the field on which  the covariant derivative applies (and in one case also crucially on the DM mass), and if this particle is not a SM singlet on the value of its millicharge.
This leads to upper bounds on the intensity of the $\gamma$-ray line produced, given in Fig. \ref{fig:Kin_mix}. This figure shows that if the DM is only charged under the dark sector, it can lead to a line matching the present experimental sensitivities without overshooting the bound on the flux of antiprotons and diffuse photons. On the contrary, when the particle emitting the photon from its millicharge is also charged under the SM, the cosmic rays constraints are much stronger than direct searches for spectral lines. Therefore, in this case, if a line were to be detected with energy above the $Z$ mass and with about the present experimental sensitivity, it could not be explained in such a way.
Such a conclusion can also hold for $m_{DM}$ far below the $Z$ mass. 

For the massless hidden gauge boson case (and the massive case where $m_{Z'}$ is both below the GeV scale and smaller than $m_{DM}$) relevant additional constraints show up imposing that the two-body decay width to
a $\gamma'$ (Z') leads to a lifetime longer than the age of the Universe. Combining this constraint with the direct detection bounds on a millicharge,  an observable $\gamma$-ray line requires small values of the dark charge $g'Q'$.

 As for a decay into a neutrino and a photon, given the stringent constraints that exist on the millicharge of a neutrino, the $Z$ mediated decay into three neutrinos, or into a neutrino and a electron-positron pair, this possibility is forbidden unless $m_{DM}$ is below the MeV scale. For lower masses, and down to the KeV scale, an observable line induced in this way is not excluded by these considerations. Such a neutrino millicharge scenario could even be at the origin of the recently reported, yet to be confirmed, 3.5 KeV X-ray line.

\vspace{5mm}
\section*{Acknowledgement}
\vspace{-3mm}
We acknowledge stimulating discussions with M.~Gustafsson and M.~Tytgat.
This work is supported by the FNRS-FRS, the IISN, the FRIA and the Belgian Science Policy, IAP VI-11. T.H. thanks the IFAE-UABarcelona theory group for its hospitality.

\appendix
\section{Ratios of  $D_\mu  \bar{\psi} \sigma^{\mu \nu}  D_\nu \psi_{DM } \phi $ (non-singlet DM)}

In the case in which $T_3^{DM}=T_3^\psi= 0$ but is not a singlet, the denominator of Eq.~(\ref{ratio_DDpsidpsi}) takes the value
\begin{eqnarray}
&D&= \frac{g^2}{16} (1-(\frac{M_W}{M_{DM}})^2) \left( \frac{M^2_{DM}-M^2_W}{M^2_W}  (1+10 (\frac{M_W}{M_{DM}})^2 +(\frac{M_W}{M_{DM}})^4)  \right. \nonumber\\ 
&+&\left. \frac{(M^2_{DM}-M^2_W)^3}{M^2_W M^4_{DM}} - 2 (1+(\frac{M_W}{M_{DM}})^2-(\frac{M_W}{M_{DM}})^4  -(\frac{M_{DM}}{M_W})^2) \right) \nonumber\\
&\times &  (n_{CR/W^+} + n_{CR/W^-}) \,.
\label{ratio_G}
\end{eqnarray}
In the case in which $T_3^{DM}=T_3^\psi \neq 0$, the denominator takes the value
\begin{eqnarray}
&D&=(1-(\frac{M_Z}{M_{DM}})^2)^2 (1+\frac{1}{2}(\frac{M_Z}{M_{DM}})^2) \frac{g^2}{4 c^2_\epsilon}  n_{CR/Z}  \nonumber\\
&+& \frac{g^2}{32} (1-(\frac{M_W}{M_{DM}})^2) \left( \frac{M^2_{DM}-M^2_W}{M^2_W}  (1+10 (\frac{M_W}{M_{DM}})^2 +(\frac{M_W}{M_{DM}})^4)  \right. \nonumber\\ 
&+&\left. \frac{(M^2_{DM}-M^2_W)^3}{M^2_W M^4_{DM}} - 2 (1+(\frac{M_W}{M_{DM}})^2-(\frac{M_W}{M_{DM}})^4  -(\frac{M_{DM}}{M_W})^2) \right) \nonumber\\
&\times &  (n_{CR/W^+} + n_{CR/W^-}) \,.
\label{ratio_H}
\end{eqnarray}

\end{document}